\documentclass[11pt,a4paper]{scrartcl}

\usepackage{CLICdp}

\usepackage{CLICdp_definitions}
\usepackage[symbol]{footmisc}
\usepackage{feynmf}
\usepackage{textpos}
\usepackage{sansmath}

\title{Measurement of the Higgs decay to electroweak bosons at low and intermediate CLIC energies}

\clicdpconf{2016}{004}  

\date{\formatdate{25}{3}{2016}}

\addauthor{I.~Bo\v{z}ovi\'{c}-Jelisav\u{c}i\'{c}\thanks{ibozovic@vinca.rs}}{\institute{1}}
\addauthor{G.~Milutinovi\'{c}-Dumbelovi\'{c}}{\institute{1}}
\addauthor{M.~Pandurovi\'{c}}{\institute{1}}
\addauthor{S.~Luki\'{c}}{\institute{1}}

\addinstitute{1}{Vinca Institute of Nuclear Sciences, University of Belgrade, Serbia}

\onbehalfof{the CLICdp collaboration}

\abstract{In this paper a simulation of measurements of the Higgs boson decay to electroweak bosons in \epem collisions at CLIC is presented. Higgs boson production and subsequent $\PH\to\zz^\ast$ and $\PH\to\PW\PW^\ast$ decay processes were simulated alongside the relevant background processes at 350~GeV and 1.4~TeV center-of-mass energy. Full detector simulation and event reconstruction were used under realistic beam conditions. The achievable statistical precision of the measured product of the Higgs production cross section and the branching ratio for the analysed decays has been determined.}

\titlecomment{Talk presented at International Workshop on Future Linear Colliders
(LCWS15), Whistler, Canada, 2-6 November 2015}

\begin{document}
\titlepage

\section{Introduction}
The future Compact Linear Collider (CLIC) offers excellent potential for a precise 
and comprehensive set of measurements of the properties of the Higgs boson. Compared 
to hadron colliders, \epem collider 
experiments offer an environment of low QCD background, fully reconstructable process 
kinematics and low radiation levels. The current staged approach to CLIC construction 
and operation foresees three stages, with 380~GeV, 1.4~TeV and 3~TeV in the 
center-of-mass (CM), with an integrated luminosity of 500~\fbinv, 1.5~\abinv and 
3~\abinv, for the three respective energy stages, in order to maximize the 
physics potential of the experiment in the shortest time. The studies presented in 
this paper have been carried-out assuming a staging scenario with 350~GeV as the CM 
energy of the first stage.

Precise exploration of the Higgs properties allows to eventually address physics 
beyond the Standard Model (SM) through deviations from the SM predicted values. 
Couplings to electroweak (EW) bosons are of particular interest to probe the Higgs 
boson structure (i.e.\ compositeness). In theoretical models extending the SM, 
the relative deviation of the Higgs couplings 
w.r.t.\ the SM is of the order $v^2/\Lambda^2$, where $v$ is the vacuum expectation 
value of the Higgs field and $\Lambda$ is the scale of the new physics. It follows, 
e.g., that a 5\% deviation can be expected for $\Lambda=1\unit{TeV}$. In this 
context, the measurement precision of the Higgs couplings determines the scale that 
can be accessed. As discussed in \cite{CLIC_snowmass13, IBJ_Lomonosov_15}, a fit to the 
full set of data from all energy stages allows extraction of all Higgs couplings with
best precision. In particular, the couplings to the vector bosons are determined with
a sub-percent uncertainty \cite{CLIC_snowmass13}. It is therefore important to optimize every single measurement for the best achievable statistical precision. 
The analyses of the measurement of the $\PH\to\zz^\ast$ and $\PH\to\PW\PW^\ast$ 
decays presented here focus on the statistical uncertainties of the measured product 
of the Higgs production cross-section and the corresponding branching fraction.

\subsection{Simulation and reconstruction}
The analyses were performed using a full simulation of the \clicild detector concept.
The \clicild concept is derived from the ILD detector proposed for the International 
Linear Collider \cite{ildloi:2009} and adapted to the CLIC experimental conditions \cite{CLIC_PhysDet_CDR}. 

Event generation for signal and background has been performed with \whizard V1.95 
\cite{whizard11,whizard2}, using \pythia V6.4 \cite{Sjostrand2006} to simulate the
hadronization processes. Initial state radiation (ISR) and a realistic beam spectrum 
are included in the simulation using results of the beam-beam simulation with 
\guineapig 1.4.4 \cite{Schulte:1999tx}. 
Equivalent photon approximation (EPA) is used to describe events with virtually exchanged 
photons below 4~GeV. Hadronic background produced from Beamstrahlung 
photons has been overlaid to the simulated events before the digitization step. 
The \pandora  \cite{thomson:pandora, Marshall2013153} algorithm is employed to ensure particle identification and energy determination. The IsolatedLeptonFinder \cite{IsolatedLeptonFinder} and TauFinder \cite{TauFinder} \marlin processors are used to identify electrons, muons and tau leptons, respectively. The LCFIplus \marlin processor was used for flavour tagging \cite{LCFIPlus}.
Unpolarised beams were assumed, so obtained results are somewhat conservative since the Higgs production cross-section can be enhanced by the beam polarization \cite{CLIC_snowmass13}.

\section{Higgs decay to a WW* pair at 350~GeV}
At the first CLIC energy stage with \roots=350 GeV, the leading Higgs production channel is the 
s-channel Higgsstrahlung process, $\epem\to\zhsm$. The Higgs decay to a W-pair is studied, 
where the W bosons decay hadronically, $\PH\to\PW\PW^{\ast}\to\qqqq$. The branching fraction 
for the decay $\PH\to\PW\PW^{\ast}$ is 21.5\% \cite{Dittmaier:2012vm}, out of which 45.6\% W-pairs
decay hadronically \cite{Dittmaier:2012vm}. Therefore the signal repersents 9.8\% of all Higgs 
boson decays. 

The final states of the \zhsm system are subdivided according to the \PZ boson decay 
into the fully hadronic final state, where the \PZ boson decays to a pair of jets 
($\PZ\to\qqbar$) and the 
semileptonic final state, where the \PZ decays to a pair of electrons or muons 
($\PZ\to\Plp\Plm; \Pl=\Pe,\PGm$). 
In Table \ref{Table1}, the list of the signal and the most relevant background processes is given with the corresponding cross sections. Higgs decays to final states other than a $\PW\PW^\ast$ pair were included as background.
                                   
\begin{table} 
\centering
\caption{\label{Table1} List of considered processes with the corresponding cross-sections at $\sqrt{s}=350 GeV$.}
\begin{tabular*}{\columnwidth}{@{\extracolsep{\fill}}ll@{}}
\hline
\multicolumn{1}{@{}l}{Signal process}  									& $\sigma(fb)$  \\
\hline
$\epem\to\zhsm,\PH\to\PW\PW^\ast\to\qqqq$       &               \\
$\PZ\to\epem$                                   & 0.453        	\\
$\PZ\to\mpmm$                                   & 0.454        	\\
$\PZ\to\qqbar$                                  & 9.16        	\\
\hline     

\multicolumn{1}{@{}l}{Background}               &               \\  
\hline 
$\epem\to\zhsm$, other Higgs decays             & 92       	\\
$\epem\to\qqqq$       				& 5850        	\\
$\epem\to\qqbar\Plp\Plm$      			& 1700        	\\
$\epem\to\qqbar\Pl\PGn$                         & 5910         	\\
$\epem\to\qqbar\nunubar$                        & 325       	\\
$\epem\to\PH\nuenuebar$			        & 52       	\\
$\epem\to\ttbar$                     		& 450        	\\
$\epem\to\wwz$		                        & 10		\\
\hline
\end{tabular*}
\end{table}

\subsection{Method}

Event selection is performed in several steps. First the event type is determined by 
searching for isolated leptons from the \PZ decay. The fully hadronic or semileptonic 
final states are identified as events containing either zero or two isolated leptons, 
respectively. In the case of a semileptonic final state, the particle flow objects 
(PFOs) that are assigned to leptons are removed from the event and the rest of the event is 
clustered into four jets using the \kT clustering algorithm \cite{S.Catani}. The hadronic 
final states are clustered into six jets. 
To each jet in an event, \PQb and \PQc-tagging probabilities are assigned.

The next step is the identification of the Higgs candidate, one on-shell and one off-shell 
\PW boson candidate and the \PZ boson candidate. For the hadronic 
final state the best combination of 
jet pairs is found by minimisation of the total $\chi^2$ of the corresponding boson masses: 

\begin{equation}
\chi^2=\frac{(m_{ij}-m_{\PW})^{2}}{\sigma^{2}_{m_{\PW}}}+\frac{(m_{kl}-m_{\PZ})^{2}}{\sigma^{2}_{m_{\PZ}}}+\frac{(m_{ijmn}-m_{\PH})^{2}}{\sigma^{2}_{m_{\PH}}}, \hspace{0.5cm} i,j,k,l,m,n=1,6
\end{equation}

\noindent where $\sigma_{m_{\text{V}}}$ (V = \PW, \PZ, \PH) stands for the measured 
widths of the corresponding boson invariant mass distributions. For the semileptonic 
final state, the \PZ boson is reconstructed using the selected leptons, while the 
on-shell \PW boson is reconstructed from the pairs of jets closest to the \PW mass:

\begin{equation}
(\Delta m_{\PW})_{\min}=\min |m_{\PW}-m_{ij}|. \hspace{2cm} i,j=1,4
\end{equation}

A preselection based on kinematic variables is applied in order to reduce large cross-section background processes. After the preselection, a multivariate analysis (MVA) \cite{Hoecker2007} event selection 
with the Boosted Decision Tree (BDT) classifier is performed on the basis of kinematic properties of the event, in order to reject the residual background.
The expected relative statistical uncertainty of the product of the Higgsstrahlung 
cross-section and the corresponding branching ratio is calculated as:

\begin{equation}
\delta(\sigma \times \br{}) = \frac{\sqrt{\ns+\nb}}{\ns},
\label{Eq.3}
\end{equation}

\noindent where \ns and \nb denote the number of selected signal and background events.

\subsection{Preselection}

Different preselection criteria are applied to different types of final states. For 
the semileptonic final states the conditions are imposed on the invariant mass of the 
\PZ boson (70 GeV < $m_{Z}$ < 110 GeV) and on the number of the particle flow object 
$N_{\text{PFO}}>20$ in order to reduce background from the high cross-section processes 
$\epem\to\qqbar\Pl\PGn$ and $\epem\to\qqbar\Plp\Plm$.

For the fully hadronic final state the following criteria were applied to reduce background 
from the $\epem\to\qqqq$ and $\epem\to\qqbar\nunubar$ processes: 
\begin{itemize}
\item the invariant mass of the \PZ boson candidate, $m_{\PZ} > 40 \unit{GeV}$;
\item jet transition probabilities, $-\log y_{12}< 2.0$, $-\log y_{23} < 2.6$, $-\log y_{34} < 3.0$, $-\log y_{45}< 3.2$, $-\log y_{56}<4.0$;
\item visible energy in the event, $E_{\text{vis}}$ > 250 GeV;
\item number of particle flow objects, $N_{\text{PFO}}>50$;
\item event thrust, $\text{thrust}<0.9$;
\item \PQb-tag probabilities of all jets, $P_{\PQb}^{\text{jet}_i} < 0.9$.
\end{itemize}

The signal preselection efficiency is 71\% for the fully hadronic final state and 80 and 
87\% for the semileptonic \epem and \mpmm final states, respectively. 

\subsection{MVA selection}

After the preselection, the MVA event selection is applied, optimized for the best statistical 
precision of $\sigma\times\br{}$ (Eq.\ (\ref{Eq.3})) with 500~\fbinv of collected data. 
The following 
observables are used for the classification of events in the analysis of all final states: 
masses of the on- and off-shell \PW boson candidates, the \PZ 
boson mass and the Higgs mass, jet transition probabilities, the transverse momentum of the Higgs 
boson jets, \PQb-tag and \PQc-tag probabilities of jets and event shape variables. 
In addition to these variables, the angle between jets that constitute \PZ and Higgs boson 
candidates are used for the selection of the hadronic final state. Similarly, the polar 
angle of the \PZ boson is used for the selection of the semileptonic final state.

\begin{figure}[h]
    \centering
    \includegraphics[width=1\textwidth]{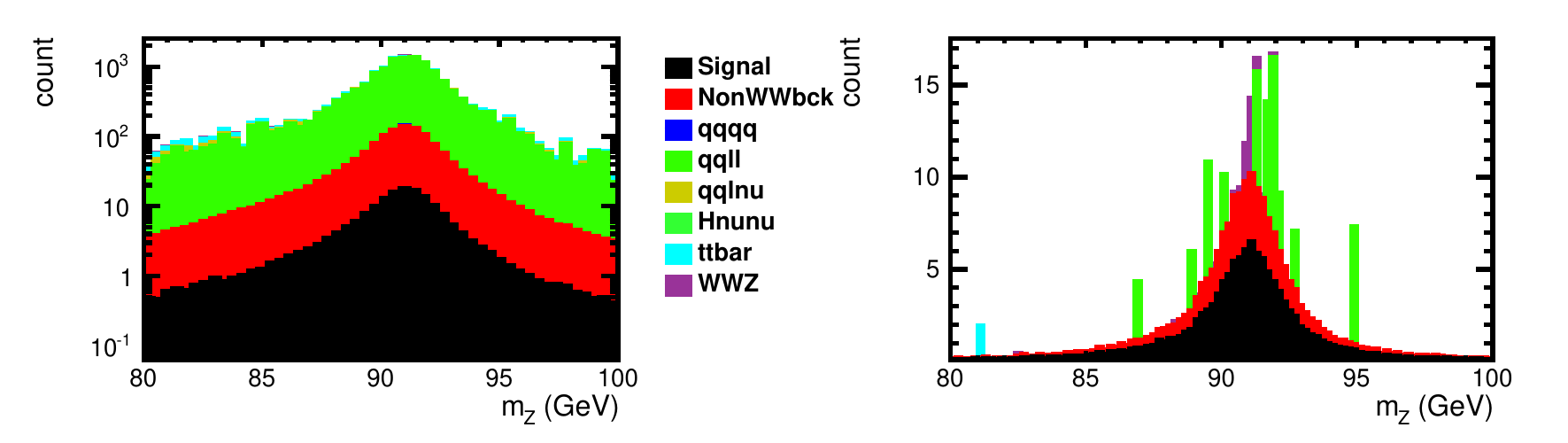}
   \begin{textblock}{1.}(10.1,-2.1)
    \bf CLICdp
    \end{textblock}
   \begin{textblock}{1.}(1.3,-2.1)
    \bf CLICdp
    \end{textblock}

    \caption{\label{Fig.1} The \PZ invariant mass distribution after the preselection (left) and after MVA selection (right) for the semileptonic $\PZ\to\mpmm$ final state, assuming 500~\fbinv of data. }

\end{figure}

\begin{figure}[h]
    \centering
    \includegraphics[width=1\textwidth]{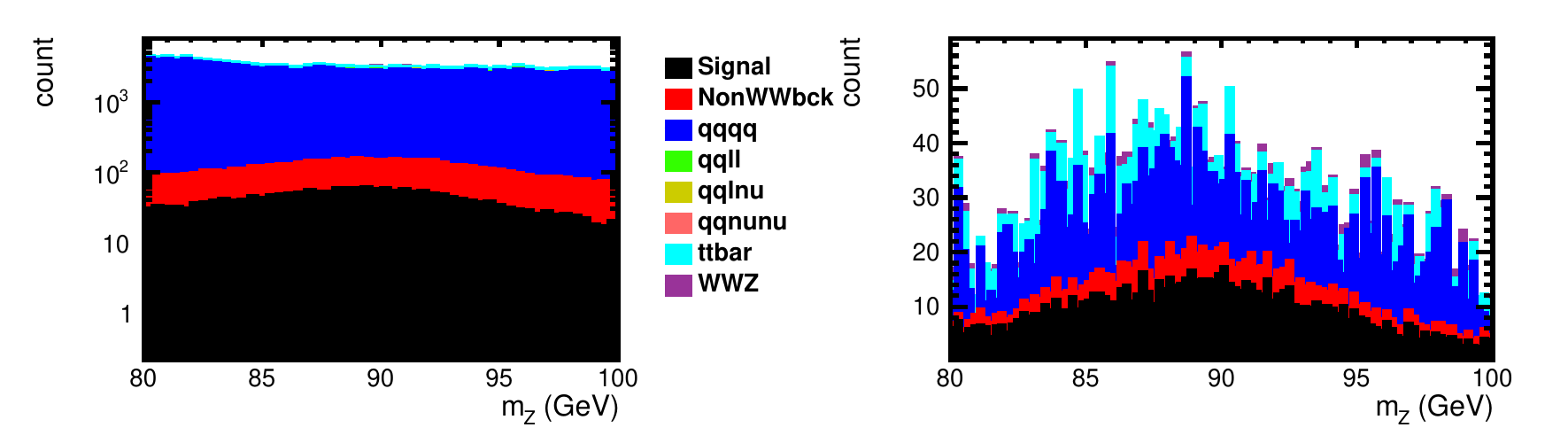}
   \begin{textblock}{1.}(1.3,-2.1)
    \bf CLICdp
    \end{textblock}

    \begin{textblock}{1.}(10.1,-2.1)
    \bf CLICdp
    \end{textblock}

    \caption{ The \PZ invariant mass distribution after the preselection (left) and after MVA 
              selection (right) for the hadronic ($\PZ\to\qqbar$) final state, assuming 500~\fbinv of data. }
    \label{Fig.2}
\end{figure}

The corresponding distributions of the Z invariant mass, after preselection and MVA event 
selection are given in Fig.\ \ref{Fig.1} and Fig.\ \ref{Fig.2}, for the semileptonic and the 
hadronic final states, respectively. After the final selection, the total signal efficiency is 
42$\%$ and 55$\%$, for the \epem and \mpmm final states, respectively. For the 
fully hadronic final state the overall signal efficiency is 29$\%$.

The expected statistical accuracy for the measurements of 
$\sigma_{\zhsm}\times\br{\PH\to\PW\PW^{\ast}}$ is found to be 
5.9\%, 13.1\% and 16.1\% for the hadronic, \mpmm and \epem final states of the 
\PZ boson, respectively. Detailed 
results for the $\sigma_{\zhsm}\times\br{\PH\to\PW\PW^{\ast}}$ measurement are listed 
in Table \ref{TabFin}.

\begin{table}[!htbp]
\centering
\caption{\label{TabFin}  Summary of the simulation results for the 
   $\sigma_{\zhsm}\times\br{\PH\to\PW\PW^{\ast}\to\qqqq}$ measurement at 
   $\sqrt{s}=350\unit{GeV}$ CLIC with unpolarised beams and 500~\fbinv of data. 
   \ns is the number of signal 
   events in the final selection, \es represents the overall signal efficiency and 
   $\delta(\sigma \times \br{})$ is the relative statistical accuracy of the 
   measured observable.}
\begin{tabular*}{\columnwidth}{@{\extracolsep{\fill}}llll@{}}
\hline
\multicolumn{4}{c}{ $\sigma_{\zhsm}\times\br{\PH\to\PW\PW^{\ast}\to\qqqq}$ }	  \\  
\hline
		  			&  \PZ$\to$\epem &  \PZ$\to$\mpmm &	\PZ$\to$\qqbar 	\\
\hline
\ns   		  	   	& 95  	  & 125	   & 1328     		\\
\es		  	    	& 42\%    & 55\%   & 29\%      	\\
$\delta(\sigma \times \br{})$   & 16.1\%    & 13.1\%   & 5.9\% 	   	\\
\hline

\end{tabular*}
\end{table}




\section{Higgs decay to a ZZ* pair at 1.4 TeV}
\label{section3}
In the WW-fusion process $\epem \to \PH \nuenuebar$ at 1.4 TeV center-of-mass 
energy, $\PH\to \zz^\ast$
decays can be measured in the fully hadronic ($\zz^\ast\to \qqbar\qqbar$) and 
in the semileptonic ($\zz^\ast\to \qqbar \Plp\Plm ;  \Pl = \Pe, \PGm,  \PGt$) 
final states. The experimental signature is 4 jets, or 2 jets and 2 leptons in the 
final state, respectively. The total invariant mass of the final state should be 
consistent with $m_{\PH}$ and the invariant mass of one pair
of jets or of the lepton pair should be consistent with  $m_{\PZ}$. 

The branching fraction for the decay $\PH\to \zz^\ast$ is 2.89\%  \cite{Dittmaier:2012vm}. 
The fully hadronic final
state of a $\zz^\ast$ pair has a branching ratio of 48.9\% \cite{Dittmaier:2012vm}, 
resulting in an effective cross section of 3.45 fb for the 
$\epem\to\PH\nuenuebar\to\zz^\ast\nuenuebar\to\qqqq\nuenuebar$ process, while
the semileptonic final state of a $\zz^\ast$ pair has a branching ratio of 14.1\% 
\cite{Dittmaier:2012vm} and the effective cross section for the 
$\epem\to\PH\nuenuebar\to\zz^\ast\nuenuebar\to\qqbar\Plp\Plm\nuenuebar$ process is
0.995 fb. This results in 5175 and 1492 signal events, respectively, for the hadronic 
and the semileptonic final states and for an integrated luminosity of 1.5~\abinv. In Table 
\ref{tableprocess}, the list of signal and the most relevant background processes is given, 
with the corresponding cross sections. The main background process, with the same final state
particles as the fully hadronic signal final state, is $\epem \to \PH \nuenuebar \to \PW\PW^\ast \nuenuebar \to \qqqq \nuenuebar$.
Other important background processes are $\Pepm\PGg\to\qqqq\PGn$, 
$\PGg\PGg\to\qqqq$ and $\Pepm\PGg\to\qqqq\Pepm$, due to their large
cross sections. The latter can be substantially reduced by requiring high-$\text p_{\text T}$ jets, 
above 80~GeV, at
the preselection level. Other background processes can be discriminated from the signal events using a multivariate approach.

\begin{table} 
\centering
\caption{\label{tableprocess} List of considered processes with the corresponding 
effective cross-sections at 1.4~TeV. }
\begin{tabular*}{\columnwidth}{@{\extracolsep{\fill}}ll@{}}
\hline
\multicolumn{1}{@{}l}{Signal processes}  & $\sigma(\text{fb})$ \\
\hline
$\epem\to\PH\nuenuebar\to\zz^\ast\nuenuebar\to\qqqq\nuenuebar$      & 3.45         \\
$\epem\to\PH\nuenuebar\to\zz^\ast\nuenuebar\to\qqbar\Plp\Plm\nuenuebar$   & 0.995         \\
\hline     
\multicolumn{1}{@{}l}{Background}  & $\sigma(\text{fb})$ \\  
\hline 
$\epem\to\PH\nuenuebar\to\PW\PW^\ast\nuenuebar\to\qqqq\nuenuebar$   & 27.6         \\ 
$\epem\to\PH\nuenuebar\to\bb\nuenuebar$                     & 137       \\
$\epem \to \qqbar\nuenuebar $                      & 788        \\
$\epem \to \qqbar \Plp\Plm$                      & 2730      \\
$\PGg\PGg \to \qqbar \Plp\Plm $                      & 13800      \\
$\epem \to \qqbar \Pl\PGn$                      & 4310      \\
$\Pepm\PGg\to \qqqq \PGn$                 & 338       \\
$\PGg\PGg\to\qqqq$                        & 30200 \\
$\Pepm\PGg\to\qqqq\Pepm$                  & 2890 \\
\hline
\end{tabular*} 
\end{table}

\subsection{Method}
For the semileptonic final state, the first step is the search for isolated leptons 
(electrons, muons or taus). Exactly two leptons are required,
otherwise the event is rejected. If two isolated leptons are found, all remaining 
particles in the event are clustered into two jets using the \kT algorithm. For the hadronic final state, the event is directly
clustered by the \kT algorithm into four jets. For both final states, flavour-tagging 
is performed to reduce the background from the dominant $\PH\to\bb$  process.

Preselection based on kinematic variables is applied as described in Sec.\ \ref{section32} 
to reduce the high cross-section background. After the preselection, an MVA
event selection based on the BDT classifier is performed to minimize the residual background. The expected
statistical accuracy on $\sigma_{\PH\nuenuebar} \times \br{\PH\to\zz^\ast}$ is calculated as in Eq.\ (\ref{Eq.3}).

\subsection{Preselection}
\label{section32}
Two leptons and two jets of the semileptonic final state, or 4 jets of the hadronic final
state are paired to form the \PZ boson candidates. Since $m_{\PH} < 2 m_{\PZ}$, 
one on-shell and one off-shell \PZ boson are produced in the Higgs decay. Thus, one \PZ candidate 
is required to have invariant mass consistent with $m_{\PZ}$ (on-shell \PZ boson), 
while the second candidate is required to form the off-shell \PZ boson. 
The preselection cuts for the fully hadronic final state are:

\begin{itemize}
\item mass of the on-shell \PZ boson: 45~GeV < $m_{\PZ}$ < 110~GeV;
\item invariant mass of the off-shell Z boson candidate: $m_{\PZ^\ast}$ < 65~GeV;
\item higgs invariant mass: 90~GeV < $m_{\PH}$ < 165~GeV;
\item jet transition probabilities: $-\log y_{34}$ < 3.5, $-\log y_{23}$ < 3.0;
\item visible energy: 100~GeV < $E_{\text {vis}}$ < 600~GeV;
\item missing transverse momentum: $\pT^{\text {miss}}$ > 80~GeV;
\item \PQb-tag probabilities of all jets: $P_{\PQb}^{\text {jet}_i}$ < 0.95.
\end{itemize}

For the semileptonic final state, the only preselection criterion is that exactly two 
isolated leptons are found.
The signal preselection efficiency is 32\% for the fully hadronic final state and 62\% for
the semileptonic final state. Only 0.8\% of the four-jet backround events 
${\Pepm\PGg\to\qqqq\PGn}$, $\PGg\PGg\to\qqqq$ and $\Pepm\PGg\to\qqqq\Pepm$ remain after 
the preselection. Yet, these background still dominate over the signal 
due to the large cross-sections (see Fig.\ \ref{qqqq1} left).

\subsection{MVA event selection}
For the fully hadronic final state, 11 sensitive observables are used for the
classification of events: masses of the on- and off-shell \PZ bosons, the Higgs mass, 
jet transition probabilities, visible energy, missing transverse momentum, \PQb-tag and \PQc-tag
probabilities of the jets. For the semileptonic final state 17 observables are used:
masses of the on- and off-shell \PZ bosons, invariant masses of the dijet and dilepton systems,
the Higgs mass, jet transition probabilities, visible energy in the event, the difference
between the visible energy in the event and the Higgs visible energy, missing transverse
momentum, \PQb-tag and \PQc-tag probabilities of the jets, polar angle of the Higgs candidate and the number of all PFOs in the event.

\begin{figure*}
\centering
\includegraphics[width=1\textwidth]{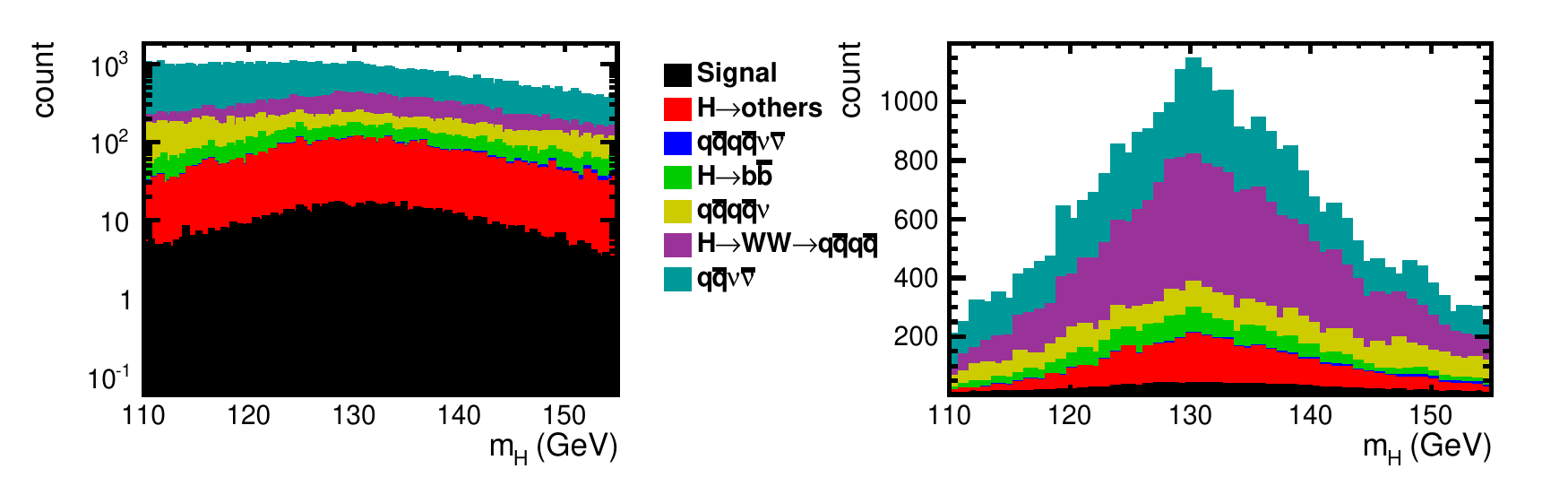}
\begin{textblock}{0.2}(3.5,-2.3)
\bf CLICdp
\end{textblock}
\begin{textblock}{0.2}(10.2,-2.3)
\bf CLICdp
\end{textblock}
\caption{\label{qqqq1} Stacked histograms of the Higgs invariant mass distributions 
with preselection only (left) and after MVA selection (right) for the fully hadronic 
final state with 1.5~\abinv of data.}
\end{figure*}

\begin{figure*}
\includegraphics[width=1\textwidth]{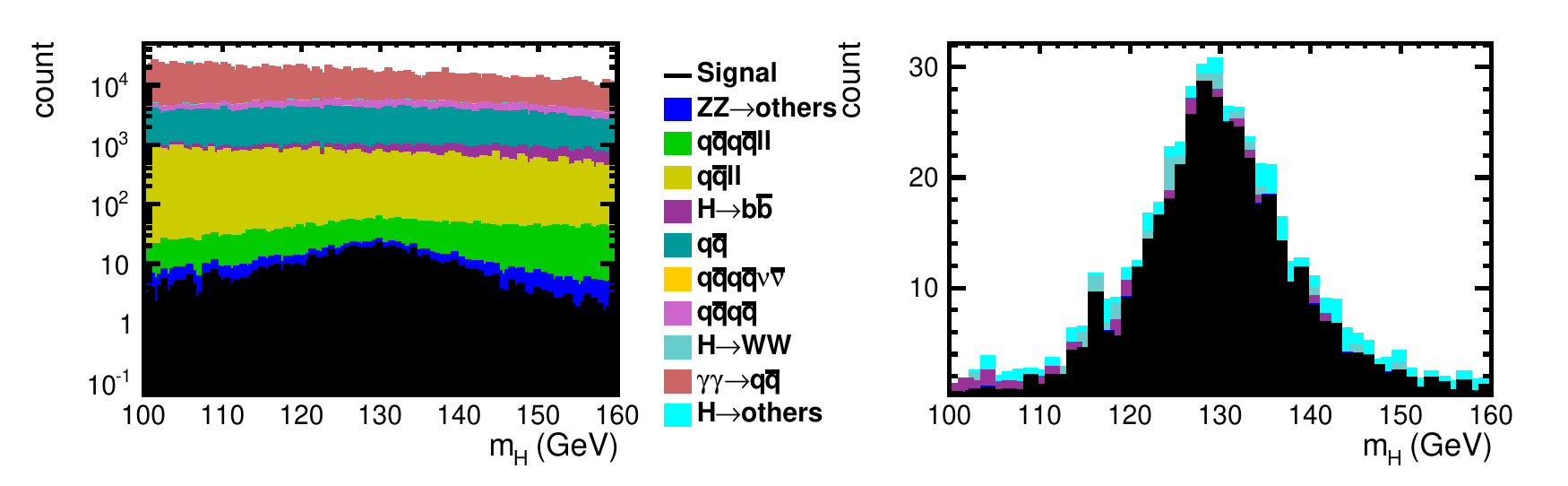}
\begin{textblock}{0.2}(3.5,-2.3)
\bf CLICdp
\end{textblock}
\begin{textblock}{0.2}(10.2,-2.3)
\bf CLICdp
\end{textblock}
\caption{\label{qqll} Stacked histograms of the Higgs invariant mass distributions 
 with preselection only (left) and after MVA selection (right) for the semileptonic 
final state with 1.5~\abinv of data.}
\end{figure*}

In both final states, the BDT cut minimizing the statistical uncertainty is chosen (Eq.\ (\ref{Eq.3})),
giving an overall efficiency of 20\% and 28\%, for the fully hadronic and semileptonic final
states, respectively. The Higgs invariant mass distribution in the fully hadronic events 
that have passed the preselection is 
given in \mbox{Fig.\ \ref{qqqq1}} (left), while Fig.\ \ref{qqqq1} (right) shows the same 
distribution for events passing the BDT selection. 
The equivalent distributions are shown for the semileptonic final state in Fig.\  \ref{qqll}.

Statistical uncertainties of $\sigma_{\PH\nuenuebar} \times \br{\PH\to\zz^\ast}$, derived from Eq. (\ref{Eq.3}), are presented in Table \ref{tablefinal} for both final states.

\begin{table}
\centering
\caption{\label{tablefinal}Summary of the simulation results for the 
$\sigma_{\PH\nuenuebar} \times \br{\PH\to\zz^\ast}$
measurement at 1.4~TeV CLIC with unpolarised beams and for an integrated luminosity
of 1.5~\abinv. \ns is the number of 
selected signal events, \es represents the overall signal efficiency and $\delta(\sigma\times \br{})$ is the relative statistical 
accuracy of the measured observable.}
\begin{tabular*}{0.7\columnwidth}{@{\extracolsep{\fill}} l r r @{}}
\hline
\multicolumn{3}{c}{ $\sigma_{\PH\nuenuebar} \times \br{\PH\to\zz^\ast}$}	  \\
\hline
                             & $\zz^\ast\to\qqqq$ & $\zz^\ast\to\qqbar\Plp\Plm$ \\
\hline
\ns                        & 1031      & 425 	       	\\
\es                 & 20\%      & 28\%       	\\
$\delta(\sigma\times\br{})$    & 17.7\%    & 5.6\%  	\\
\hline
\end{tabular*}
\end{table}

\section{Summary}
Detailed analyses of Higgs decays to EW bosons are simulated at low and intermediate CLIC energies. Each analysis is optimized for the best statistical uncertainty. 

The product of the Higgsstrahlung cross section and the branching ratio for the decay to a pair 
of \PW bosons, $\sigma_{\PH\PZ} \times \br{\PH\to\PW\PW^{\ast}}$, is measured at the 350~GeV CM energy, 
by counting events with fully hadronic decays of \PW and hadronic and leptonic \PZ decays. 
The relative statistical uncertainties of the signal count, \ns, are 5.9\% for the hadronic \PZ final state and 13.1\%, 16.1\% from \PZ decays into muons or electrons, respectively.

The product of the Higgs production cross section in \ww fusion and the branching ratio for the decay to a pair of \PZ bosons, $\sigma_{\PH\nuenuebar} \times \br{\PH\to\zz^{\ast}}$, is measured at 1.4~TeV CM energy by counting hadronic and semileptonic \PZ decays. The relative statistical uncertainties of the signal count, \ns, are 5.6\% and 17.7\% for the semileptonic and hadronic final states, respectively. 

In both analyses, the statistical uncertainty is dominated by the irreducible background processes and the limited number of signal events in the experiment.

The results of the measurements presented here form part of the complete set of data from all CLIC energy stages used in a global fit to enable determination of the Higgs couplings with the 
ultimate precision. The couplings to the EW bosons are obtained at a percent or sub-percent level from the model-independent and the model-dependent fits, respectively \cite{higgspaper, IBJ_Lomonosov_15}.

\printbibliography[title=References]

\end{document}